# СУПУТНИКОВИЙ ТЕЛЕСКОП ЕЛЕКТРОНІВ І ПРОТОНІВ СТЕП-Ф НАУКОВОГО КОСМІЧНОГО ПРОЕКТУ «КОРОНАС-ФОТОН»

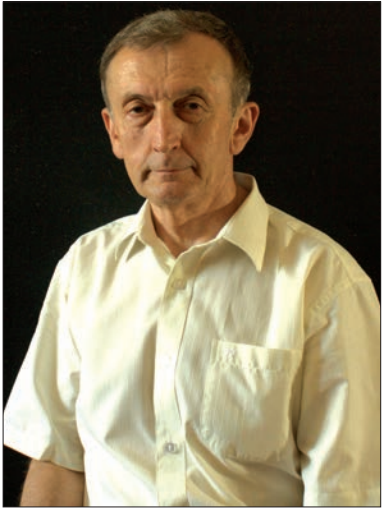

**ДУДНИК**
**Олексій Володимирович —**
доктор фізико-математичних наук, провідний науковий співробітник відділу космічної радіофізики Радіоастрономічного інституту НАН України

http://orcid.org/0000-0002-5127-5843

*Український супутниковий телескоп електронів і протонів СТЕП-Ф у 2009 р. здійснив науковий експеримент у космосі на борту низькоорбітального супутника «КОРОНАС-Фотон». У статті розглянуто передумови його створення, етапи розроблення, виготовлення і тестування габаритно-вагового, лабораторного, технологічного, льотного та запасного зразків приладу, а також комплектів контрольно-випробувальної апаратури до нього. Наведено принципи роботи, конструкцію, технічні та наукові характеристики приладу; етапи його настройки, градуювання і здійснення автономних, стикувальних, комплексних, передстартових та льотних випробувань. Наукові результати, отримані з приладу СТЕП-Ф у період глибокого мінімуму сонячної активності, дозволили встановити нові особливості в розподілах високоенергетичних заряджених частинок у радіаційних поясах Землі, в зоні Бразильської магнітної аномалії та поза їх межами.*

***Ключові слова:** магнітосфера Землі, супутниковий прилад, сцинтиляційний детектор, кремнієва матриця, заряджені частинки, радіаційний пояс, сонячна активність, Бразильська магнітна аномалія.*

У 2005—2009 рр. на замовлення Національного космічного агентства України (нині — Державне космічне агентство України) було розроблено спектрометр-телескоп заряджених частинок високих енергій СТЕП-Ф з метою вивчення поведінки космічної радіації високої енергії на супутникових висотах. Харківський національний університет ім. В.Н. Каразіна разом із Науково-дослідним інститутом радіотехнічних вимірювань (НДІРВ), Інститутом сцинтиляційних матеріалів (ІСМА) і Науково-дослідним інститутом мікроприладів НТК «Інститут монокристалів» НАН України виконали проектування, розроблення робочої конструкторської та експлуатаційної документації, виготовили і провели автономні, стикувальні, комплексні, передстартові та льотні випробування приладу СТЕП-Ф на підприємствах ракетно-космічної галузі і в наукових організаціях України і Росії та на борту низькоорбітального космічного апарата «КОРОНАС-Фотон».

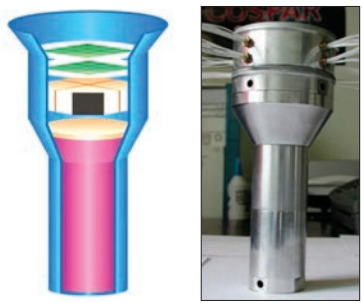

**Рис. 1.** Схематичне зображення і зовнішній вигляд детекторної головки блока детекторів СТЕП-ФД

Поштовхом для розроблення космічного приладу зі значними геометричним фактором і чутливістю до реєстрації надслабких потоків частинок було схвалення його концепції та включення до складу наукової апаратури українського супутникового проекту з пошуку провісників сейсмічної активності «Попередження». Виконавши низку робіт з ескізного проектування, наприкінці 1990-х років вдалося побудувати концепцію приладу, сформувати колектив науковців і інженерів, визначити організації — розробники детекторів, механічних вузлів та електронних модулів з максимальним використанням сенсорів, радіокомпонентів та інших складових вітчизняного виробництва.

Ініціаторам створення приладу СТЕП вдалося зацікавити фахівців Національного агентства з розвитку космічних досліджень Японії NASDA (зараз — Японське аерокосмічне агентство JAXA), і вони, за посередництва Українського науково-технологічного центру (УНТЦ), підтримали цей проект, надавши розробникам регулярний грант № 1578, фінансований урядом Японії.

Згодом стало очевидно, що з ряду причин супутниковий проект «Попередження» реалізувати не вдасться. Тоді, з огляду на значний доробок і великий обсяг уже виконаних робіт з розроблення приладу, за ініціативою харківських учених концепцію приладу було представлено керівництву проекту зі створення комплексу наукової апаратури «Фотон» наукового космічного проекту «КОРОНАС-Фотон» і після детального розгляду концепцію було схвалено. Український прилад включили до складу наукової апаратури, яку розробляли спеціалісти Росії, Польщі та Індії для вивчення жорсткого випромінювання Сонця, сонячноземних взаємозв'язків; для моніторингу потоків електронів, протонів і α-частинок високих енергій у радіаційних поясах Землі та поза ними в періоди сонячної та геомагнітної активності; для вивчення динаміки сонячних космічних променів і просторово-часових розподілів частинок високих енергій на орбітальних висотах.

Наприкінці січня 2009 р. з північного космодрому «Плесецьк» українська ракета-носій «Циклон-3» вивела супутник «КОРОНАС-Фотон» на колову орбіту висотою близько 550 км і нахилом 84°. Прилад СТЕП-Ф працював у моніторинговому режимі впродовж усього періоду активного функціонування космічного апарата — до грудня 2009 р. і зібрав інформацію про динаміку потоків частинок у магнітосфері Землі і поза її межами в рік аномально низької сонячної активності.

## Принцип роботи, технічні ідеї і конструкція приладу СТЕП-Ф

Прилад СТЕП-Ф складався з двох блоків: блока детектування СТЕП-ФД, розміщеного ззовні космічного апарата, і блока електроніки СТЕП-ФЕ, встановленого всередині герметичного відсіку з аргонно-повітряною сумішшю під атмосферним тиском. Основу блока детектування становила детекторна головка з сенсорами вітчизняного виробництва — позиційно-чутливими кремнієвими матричними і сцинтиляційними детекторами заряджених частинок (рис. 1). Ледь помітні сигнали від детекторів уловлювали 69 підсилювачів, які формували однакові за формою і тривалістю імпульси для подальшої обробки у цифровому вигляді блоком електроніки СТЕП-ФЕ. Апаратура цифрової обробки виконувала функції з визначення сортів і енергій частинок за класичною в ядерній фізиці $\Delta E$—$E$-методикою ($E$ — енергія зарядженої частинки). Як «тонкий» пролітний детектор, що формує імпульс, пропорційний $\Delta E$, було задіяно верхню кремнієву матрицю. Роль «товстого» детектора повного поглинан-ня, тобто такого, що поглинає решту енергії частинки і формує імпульс, пропорційний енергії частинки $E$, виконував шар з 4 однакових сцинтиляційних детекторів з монокристалів CsI(Tl) і кремнієвих фотодіодів з активною площею 100 мм$^2$ (10×10 мм) кожний.

Інформаційні масиви з науковими даними формувалися з періодичністю 30 с і містили дані про потоки, сорти і кути надходження частинок з мінімальною часовою роздільною здатністю 2 с. Керування приладом у польоті відбувалося за допомогою функціональних команд, які фахівці харківської лабораторії формували щодня на основі експрес-аналізу працездатності приладу. Команди в узгодженому заздалегідь і перевіреному на надійність форматі через ftp-протокол передавали до головної організації з комплексу наукової апаратури «Фотон» — Інституту астрофізики Московського інженерно-фізичного інституту (ІАФ МІФІ).

У процесі розроблення та виготовлення приладу СТЕП-Ф було використано низку нових технічних ідей:

• розроблено і виготовлено для застосування в космосі кремнієві матричні детектори, які надають приладу чутливість до напряму приходу первинних заряджених частинок, а також підвищують співвідношення корисних сигналів до електричних і теплових шумів (рис. 2);

• розроблено і виготовлено «товсті» сцинтиляційні детектори пентагональної форми на основі монокристалів йодиду цезію, активованого талієм (CsI(Tl)), з підвищеним світловим виходом, які дають змогу отримувати енергетичні спектри електронів, протонів і α-частинок у різних прошарках магнітосфери Землі (рис. 3);

• як реєстратори слабких світлових спалахів у монокристалах CsI(Tl) застосовано кремнієві фотодіоди великої площі замість традиційних вакуумних скляних фотоелектронних помножувачів, що дозволяє зменшити розміри і вагу детекторного модуля та знизити його енергоспоживання і тепловиділення;

• застосовано онлайн-обробку сигналів від кожної частинки з визначенням сорту, енергії,

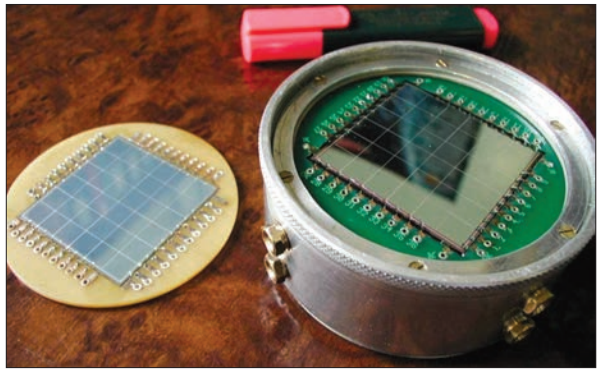

**Рис. 2.** Позиційно чутливі кремнієві матричні детектори блока детекторів СТЕП-ФД

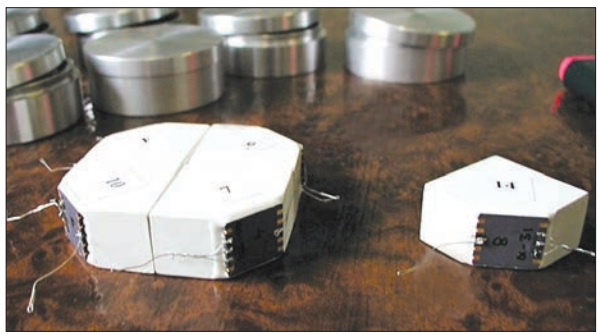

**Рис. 3.** Сцинтиляційні детектори на основі монокристалів CsI(Tl) і кремнієвих фотодіодів блока детекторів СТЕП-ФД

часу, просторового розташування і кількості частинок за допомогою мікропроцесорів і програмованих логічних інтегральних схем (ПЛІС) у блоці СТЕП-ФЕ на основі попереднього комп'ютерного моделювання методом Монте-Карло процесів проходження частинок крізь речовину детекторів з використанням ЦЕРНівської бібліотеки програм GEANT4;

• розроблено і впроваджено отримання наукових даних та інформації про працездатність приладу, формування команд керування приладом під час льотних випробувань безпосередньо з харківської лабораторії і забезпечення зв'язку з Центром зберігання даних в ІАФ МІФІ;

• забезпечено проведення експрес-аналізу наукових даних за допомогою спеціального

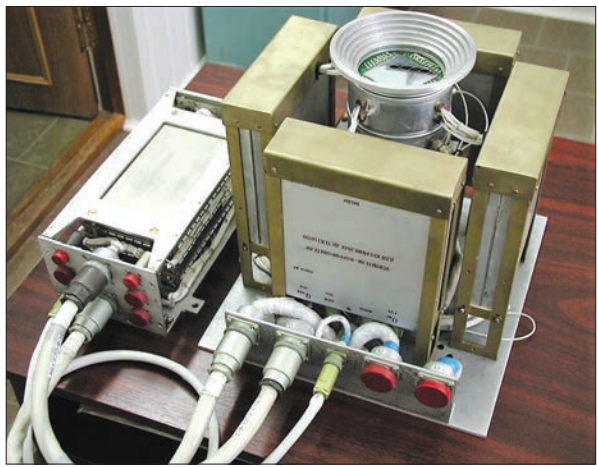

***Рис. 4.*** Лабораторний зразок приладу СТЕП-Ф у відкритому вигляді

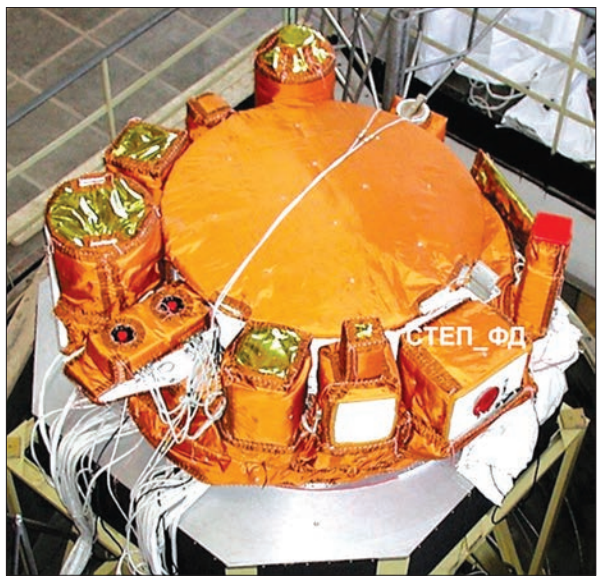

***Рис. 5.*** Блок детекторів СТЕП-ФД технологічного зразка приладу СТЕП-Ф у складі зовнішніх блоків детектування комплексу наукової апаратури «Фотон» програмного забезпечення з метою оперативного керування режимами роботи приладу під час його льотної експлуатації.

Крім того, під час створення контрольно-випробувальної апаратури для технологічного і льотного зразків приладу СТЕП-Ф було розроблено програми і методики тестувань, випробувань і градуювальних вимірювань електричних характеристик та інших параметрів детекторів заряджених частинок, модулів аналогової і цифрової обробки сигналів; створено програмне забезпечення для отримання, перетворення і експрес-аналізу наукової і статусної телеметричної інформації; виготовлено імітатори службових і допоміжних систем космічного апарата тощо.

## Настройка, градуювання і випробування зразків приладу

Настройку електричних параметрів лабораторного, технологічного і льотного зразків приладу СТЕП-Ф (рис. 4, 5) здійснювали в лабораторних умовах за допомогою спеціально розробленої контрольно-випробувальної апаратури і радіоактивних ізотопів із заздалегідь обраними енергіями β-частинок і γ-квантів. Контрольно-випробувальна апаратура складалася з кількох автономних електронних блоків, імітаторів системи збирання і реєстрації наукової інформації і бортового блока керування і синхронізації. Ця апаратура дала змогу не лише відрегулювати 69 каналів аналогової обробки сигналів зовнішнього блока СТЕП-ФД, що надходили від детекторів заряджених частинок, а й перевірити правильність формування вихідних інформаційних масивів наукової інформації у блоці електроніки СТЕП-ФЕ з різними інтерфейсами зв'язку з бортовими системами космічного апарата.

Наземне градуювання спектрометра-телескопа елементарних заряджених частинок було здійснено на кількох прискорювачах частинок високих енергій, зокрема на іонному циклотроні Інституту фізичних і хімічних проблем RIKEN (Відділення інституту в Токіо). У серії експериментів, які заздалегідь замовили і профінансували японські партнери з Центру космічних досліджень у м. Цукуба Національного агентства з розвитку космічних досліджень Японії NASDA, використовували α-частинки з енергіями 400 МеВ і нестійкі іони водню з енергією 140 МеВ, частина з яких розпадалися на шляху до мішені і утворювали пари протонів з енергією 70 МеВ. Для прикладу на рис. 6 наведено результати моделювання поглинених енергій у детекторних шарах телескопічної системи детекторної головки приладу порівняно з результатами вимірювань відгуків у сцинтиляційному детекторі з монокристалів CsI(Tl) шару D3 в одиницях енергії і в амплітудах сигналів на виході підсилювача-формувача каналу детекторного шару D3 на проходження α-частинок різних енергій.

Для менших енергій частинок градуювання детекторних каналів приладу СТЕП-Ф було проведено на іонному циклотроні НДІ ядерної фізики ім. Д.В. Скобєльцина Московського державного університету ім. М.В. Ломоносова, де було задіяно пучки дейтронів з енергіями 15,3 МеВ і α-частинок з енергіями 30 МеВ, а також на електронному мікротроні Інституту електронної фізики НАН України (Ужгород) з використанням пучків електронів у широкому діапазоні енергій 2—17 МеВ.

Налагодження працездатності зразків приладу СТЕП-Ф і виконання програм комплексного відпрацювання та забезпечення надійності потребувало проведення серії випробувань за заздалегідь розробленими спеціалізованими програмами випробувань окремих вузлів і блоків. Зокрема, автономні випробування лабораторного, технологічного і льотного зразків приладу здійснювалися як в організації — розробнику приладу, так і на сертифікованій випробувальній базі харківських підприємств космічної галузі України — в НДІ радіотехнічних вимірювань та НВП «Хартрон-Сигма», де було проведено температурні, вакуумні, кліматичні, механічні та акустичні випробування.

Вхідні, стикувальні і комплексні випробування технологічного і льотного зразків СТЕП-Ф у складі комплексу наукової апаратури «Фотон» було проведено на імітаторі космічного апарата в ІАФ МІФІ (рис. 7), а також у складі власне космічного апарата «КОРОНАС-Фотон» в організації — розробнику супутника, в НДІ електромеханіки (м. Істра Московської обл.).

Нарешті, передстартові комплексні випробування космічного апарата було здійснено безпосередньо на космодромі «Плесецьк»

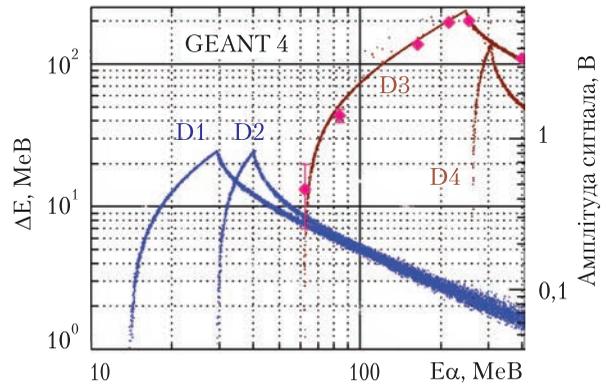

**Рис. 6.** Порівняння результатів моделювання (суцільні криві) відгуків детекторів на дію α-частинок і калібрувальних вимірювань (точки) на циклотронному прискорювачі RIKEN (Японія)

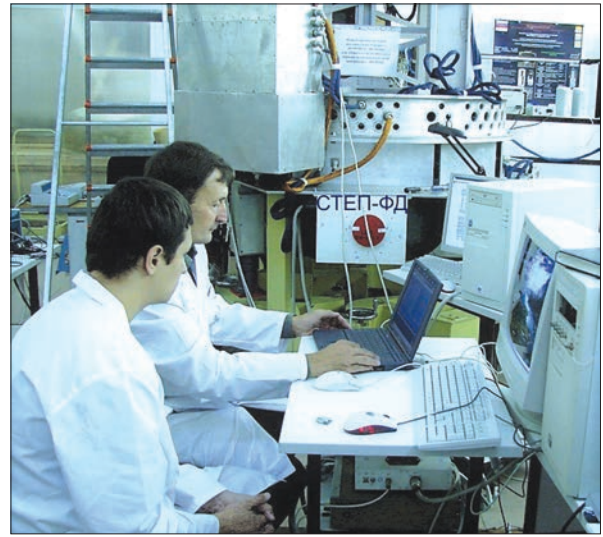

**Рис. 7.** Стикувальні випробування технологічного зразка приладу СТЕП-Ф на імітаторі космічного апарата в Інституті астрофізики Московського інженерно-фізичного інституту

(м. Мирний Архангельської обл.) (рис. 8). Під час комплексних випробувань перевіряли відпрацювання функціональних команд, наявність, внутрішню структуру, послідовність і регулярність передачі наукових і статусних цифрових масивів від блоку обробки цифрових даних СТЕП-ФЕ до системи збирання і

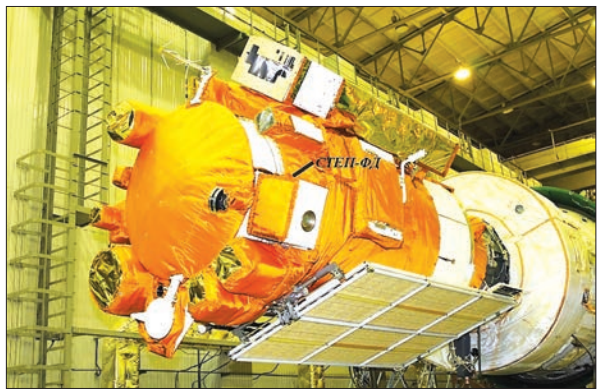

**Рис. 8.** Космічний комплекс «КОРОНАС-Фотон» перед накатуванням обтічника в монтажно-випробувальному корпусі космодрому «Плесецьк». На передньому плані космічного апарата — зовнішній блок детекторів СТЕП-ФД приладу СТЕП-Ф

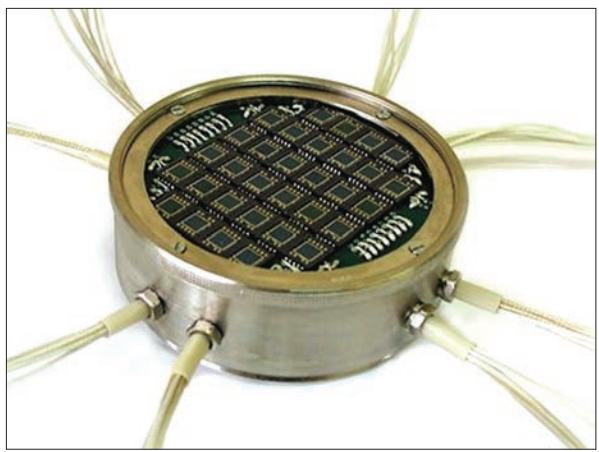

**Рис. 9.** Модуль позиційно-чутливих сенсорних шарів з фотодіодів виробництва компанії Hamamatsu Photonics для запасного комплекту блока детекторів СТЕП-ФД реєстрації наукової інформації і штатної телеметричної системи космічного апарата. Крім того, перевіряли відгук приладу на дію високоенергетичних електронів у діапазоні енергій до $E_e = 2{,}2$ МеВ від радіоактивного β-джерела $^{90}Sr$ на різних відстанях від детекторної головки.

Після завершення всіх видів наземних випробувань, що тривали протягом 2006—2009 рр., комплекс наукової апаратури космічного комплексу «КОРОНАС-Фотон», у складі якого був український супутниковий телескоп електронів і протонів СТЕП-Ф, допустили до проведення льотних випробувань. З метою експрес-аналізу якості отримуваної інформації, перетворення бінарних даних у зручний для використання і візуалізації формат було розроблено комп'ютерну програму, яка дозволяла швидко побудувати зображення залежності потоків кожного сорту частинок у кожному енергетичному інтервалі від часу з кроком 2 та 30 с.

Стандарти ракетно-космічної техніки передбачають наявність запасного комплекту приладу. Однак через недофінансування робіт з боку генерального замовника було вирішено виготовити для запасного комплекту лише найбільш критичний вузол — блок з двох позиційно-чутливих сенсорів D1 і D2 заряджених частинок блока детекторів СТЕП-ФД. За сприяння першого проректора з наукової роботи ХНУ ім. В.Н. Каразіна, члена-кореспондента НАН України І.І. Залюбовського, який від самого початку активно підтримував ідею космічного експерименту і процес створення супутникового приладу, коштом ХНУ було закуплено партію кремнієвих фотодіодів відомої японської фірми Hamamatsu Photonics, розроблено конструкторську документацію запасного комплекту сенсорів і виготовлено блок детекторних шарів D1 і D2 блока детекторів СТЕП-ФД (рис. 9).

## Типи та енергії частинок, які реєструвалися приладом СТЕП-Ф

У процесі наземних градуювальних вимірювань та попереднього аналізу перших наукових даних під час льотних випробувань, а також зіставленням з моделями розподілу зарядженої радіації на висоті 600 км було уточнено діапазони енергій, які реєструвалися приладом СТЕП-Ф (див. табл.).

Прилад визначав конкретний тип частинки, якщо вона долала щонайменше два перших детектори телескопічної системи. Частинки з невеликою енергією застрягали в першому з боку захисної фольги кремнієвому матричному де-текторі, і визначити їх сорт було неможливо. У цьому разі прилад збирав інформацію про змішану компоненту частинок — електронів і протонів чи протонів і α-частинок.

Оцінку якості наукової інформації, отриманої з приладу СТЕП-Ф на початковому етапі льотних випробувань, провели нанесенням на умовну карту Землі потоків протонів у магнітно-спокійний період з 3 березня по 1 квітня 2009 р. Складену в такий спосіб карту зіставили з результатами радіаційного картографування навколоземного простору аналогічними приладами на інших супутниках. Усереднені дані з приладу СТЕП-Ф добре узгоджувалися з даними попередніх вимірювань у періоди мінімуму сонячної активності.

**Діапазони енергій та геометричні фактори приладу СТЕП-Ф для каналів змішаної реєстрації частинок, електронів, протонів і α-частинок**

| Енергетичний діапазон, МеВ | | Геометричний фактор, см²×стер |
|---|---|---|
| *Канали змішаної реєстрації* | | |
| електрони (0,18—0,51) + протони (3,5—3,7) | | 21,7 |
| протони (3,7—7,4) + електрони (0,55—0,95) | | 21,7 |
| α-частинки (15,9—29,8) + протони (7,4—10,0) | | 21,7 |
| *Електрони* | | |
| 0,35—0,95 | | 19,5 |
| 1,2—2,3 | | 17,4 |
| ≥2,3 | | 16,2 |
| *Протони і α-частинки* | | |
| Протони | α-частинки | 19,5 |
| 7,4—10,0 | 29,8—40,5 | 17,4 |
| 15,6—17,5 | 63,8—65,2 | 16,2 |
| 17,5—19,6 | 65,2—68,5 | 15,9 |
| 19,6—22,2 | 68,5—75,4 | 15,5 |
| 22,2—25,4 | 75,4—82,0 | 15,0 |
| 25,5—29,3 | 82,0—90,0 | 14,5 |
| 29,3—33,2 | 90,0—100,0 | 13,9 |
| 33,2—38,9 | 100,0—112,0 | 13,2 |
| 38,9—46,5 | 112,0—132,0 | 12,6 |
| 46,5—55,2 | 132,0—160,0 | 12,4 |
| ≥55,2 | ≥160,0 | 19,5 |

## Наукові результати, отримані за допомогою приладу СТЕП-Ф

У процесі роботи спектрометра-телескопа СТЕП-Ф у сонячно- і геомагнітно-спокійний період 2009 р. на коловій орбіті висотою 550 км було накопичено великий обсяг наукової інформації, аналіз якої здійснювався як безпосередньо під час польоту супутника, так і після завершення експерименту. Продовжується він і нині, зокрема тривають роботи із зіставлення з даними інших приладів, встановлених на супутнику «КОРОНАС-Фотон» та на інших космічних апаратах. Під час 96-хвилинного обертання супутника навколо Землі кут між віссю поля зору приладу і найкоротшою лінією, що з'єднувала супутник та поверхню Землі, в кожний момент часу змінювався. Як наслідок, прилад СТЕП-Ф поперемінно фіксував потоки частинок, спрямованих то в бік планети, то від неї, то (двічі за 96-хвилинний оберт) вздовж поверхні Землі.

**Високоенергетичні частинки поза межами радіаційних поясів Землі.** Завдяки великій активній площі кремнієвих матричних детекторів, широкому тілесному куту зору детекторної головки приладу було підтверджено значні потоки частинок у радіаційних поясах Ван Алена, в зоні Південно-Атлантичної магнітної аномалії. Крім того, було виявлено потоки електронів проміжних енергій на низьких широтах і навіть поблизу екватора, тобто в тих областях, де, як вважалося, їх не існує. Для прикладу на рис. 10 показано наявність стійких потоків субрелятивістських електронів всюди у плазмосфері Землі, в тому числі на низьких широтах і в приекваторіальних зонах.

Під час аналізу записів з приладу СТЕП-Ф використовували результати вимірювань потоків електронів у міжпланетному просторі і на геостаціонарній орбіті, виконаних супутниками космічної погоди SOHO, ACE, STEREO-A, STEREO-B і GOES, а також сонячного вітру — з супутників ACE, WIND і SOHO. Крім того, було задіяно поширені в інтернет-мережі відкриті наземні дані про стан магнітного поля

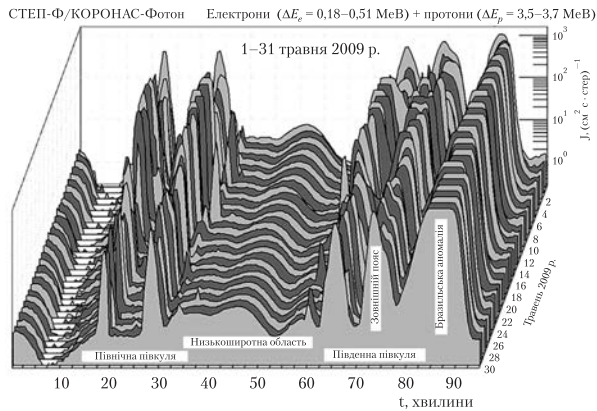

**Рис. 10.** Часовий хід потоків частинок каналу реєстрації електронів з енергіями $\Delta E_e = 0{,}18{-}0{,}51$ МеВ і протонів з енергіями $\Delta E_p = 3{,}5{-}3{,}7$ МеВ за даними СТЕП-Ф у період з 1 по 31 травня 2009 р. на першому з початку доби оберті супутника

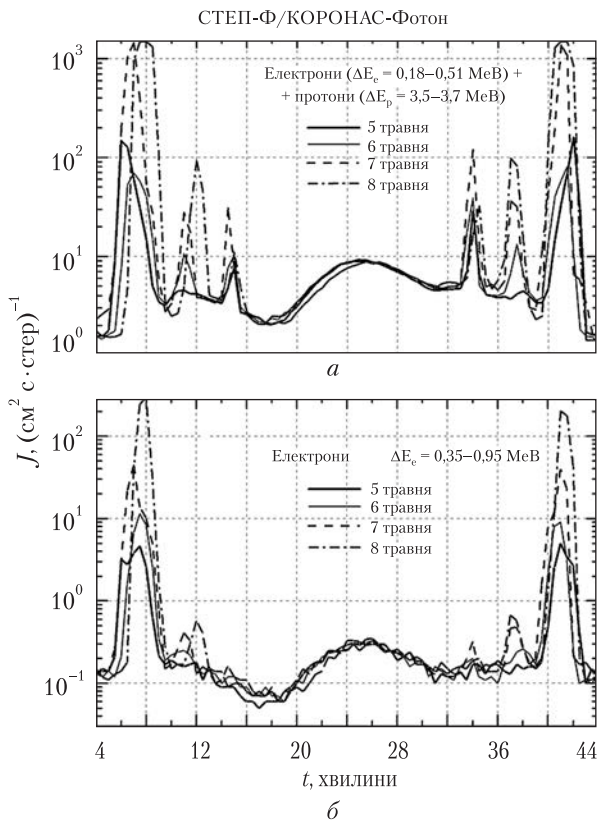

**Рис. 11.** Часовий хід інтенсивності частинок на 4-й — 44-й хв висхідних вузлів супутника у двох енергетичних каналах та іоносфери Землі. Зіставлення з різними фазами магнітних бур дало цікаві результати. Зокрема, було знайдено нову несподівану особливість, яка полягає в тому, що на початковій фазі геомагнітної бурі збільшення потоків електронів спочатку спостерігалося у внутрішньому електронному радіаційному поясі, тоді як у зовнішньому поясі зміни відбулися дещо пізніше. Під час активної фази магнітної бурі потоки електронів на зовнішній кромці зовнішнього радіаційного поясу, за даними приладу СТЕП-Ф, зросли більш ніж у 10 разів, що підтвердило результати вимірювань потоків електронів з енергіями $E_e > 2$ МеВ на геостаціонарній орбіті, проведених американськими супутниками космічної погоди серії GOES.

***Виявлення додаткового внутрішнього радіаційного поясу Землі.*** Результати вимірювань з використанням приладу СТЕП-Ф дозволили констатувати наявність додаткового внутрішнього електронного радіаційного поясу Землі, проекція якого розташована на широтах, менших за ті, на яких спостерігається проекція внутрішнього поясу Ван Алена на висоті 550 км. На рис. 11 наведено часовий хід інтенсивності частинок на 4-й — 44-й хв від початку висхідних вузлів орбіти космічного апарата «КОРОНАС-Фотон» протягом 5—8 травня 2009 р. в каналі реєстрації електронів з енергіями $\Delta E_e = 0{,}18{-}0{,}51$ МеВ і протонів з енергіями $\Delta E_p = 3{,}5{-}3{,}7$ МеВ та в каналі реєстрації електронів з енергіями $\Delta E_e = 0{,}35{-}0{,}95$ МеВ.

Третій радіаційний пояс спостерігався приладом в обох півкулях Землі; достовірність його детектування на висоті 550 км та інтенсивність електронів у ньому майже не залежали від рівня геомагнітного збурення: навіть під час мінімальної активності земного магнітного поля на межі її відсутності пояс реєструвався приладом СТЕП-Ф у низькоенергетичному діапазоні електронів. У всіх випадках обидва внутрішні пояси реєструвалися на довготах, які не збігаються з довготами розташування Південно-Атлантичної магнітної аномалії.

На рис. 12 продемонстровано часовий хід інтенсивності електронів з енергіями $\Delta E_e =$

= 0,18—0,51 МеВ на 0-й — 48-й хв від початку висхідних вузлів орбіти космічного апарата «КОРОНАС-Фотон» за період з 1 по 31 травня 2009 р. Збільшення потоків частинок у внутрішньому радіаційному електронному поясі Ван Алена і в додатковому поясі відбувається на різних магнітних оболонках, про що свідчили різні величини відношень напруженості магнітного поля $B/B_0$ в момент спостереження максимумів потоків частинок в обох поясах.

***Зіставлення зі спостереженнями частинок сонячним рентгенівським фотометром SphinX.*** Крім автономної обробки інформації з приладу СТЕП-Ф із залученням відкритих супутникових і наземних даних про космічну погоду було проведено спільний аналіз експериментальних даних, отриманих за допомогою телескопа СТЕП-Ф і польського сонячного рентгенівського фотометра SphinX про потоки заряджених частинок у навколоземному просторі. Спектрофотометр SphinX також був складовою комплексу наукової апаратури на супутнику «КОРОНАС-Фотон», але його кут зору мав ортогональну направленість до кута зору приладу СТЕП-Ф. Завдяки такому взаємному розташуванню було виявлено, що електрони в зоні Південно-Атлантичної магнітної аномалії мають анізотропний характер руху, тоді як у зовнішньому і внутрішньому радіаційних поясах Ван Алена неподалік дзеркальних точок відбиття напряму руху потоки електронів мають цілком спрямований характер.

Аналізуючи дані своїх спостережень, розробники приладів СТЕП-Ф і SphinX зробили важливий висновок про те, що під час активної фази геомагнітної бурі електрони низьких енергій (від кількох кеВ до кількох десятків кеВ) висипаються в атмосферу Землі не лише в зоні глобальної магнітної аномалії чи у відрогах радіаційних поясів, а й на всіх широтах, зокрема приекваторіальних, тобто там, де, як раніше вважалося, потоків частинок немає взагалі. В цілому, канали реєстрації рентгенівського спектрофотометра SphinX доповнили і розширили енергетичний діапазон реєстрації

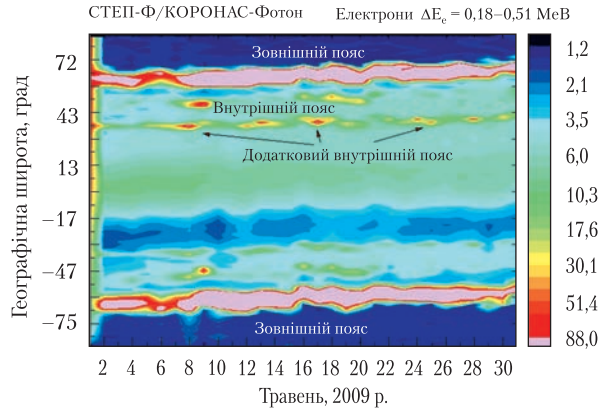

**Рис. 12.** Часовий хід інтенсивності електронів з енергіями $\Delta E_e = 0{,}18$—$0{,}51$ МеВ на 0-й — 48-й хв висхідних вузлів орбіти супутника протягом 1—31 травня 2009 р.

каналів приладу СТЕП-Ф у бік менших енергій поза зоною Південно-Атлантичної магнітної аномалії.

***Визначення емпіричних значень часу життя електронів у поясах Ван Алена.*** Мала кількість слабких спалахів на Сонці у 2009 р. і наявність ізольованих геомагнітних суббур, що не накладаються в часі одна на одну, дозволили визначити час життя підвищених та прискорених потоків електронів, які генеруються на фазах відновлення бур. Було знайдено емпіричні значення часу життя електронів зовнішнього і внутрішнього радіаційних поясів Землі за відгуками потоків частинок у максимумах їх розподілів по L-оболонках на дію двох геомагнітних збурень (рис. 13), де L — відстань від поверхні Землі на геомагнітному екваторі в радіусах Землі. Виявилося, що у зовнішньому поясі час життя електронів з енергіями $\Delta E_e = 0{,}18$—$0{,}51$ МеВ має значення $\tau_1 \approx 4{,}1 \pm 1{,}1$ доби, тоді як у внутрішньому поясі цей параметр має значення $\tau_2 \approx 1{,}5 \pm 0{,}8$ доби. Отже, відновлення стаціонарних потоків у внутрішньому поясі відбувається втричі швидше, ніж у зовнішньому поясі Ван Алена. Крім того, простежується чітка залежність часів життя електронів у цьому поясі від їх енергії — чим вища енергія частинок, тим скоріше відбувається перехід до стаціонарного стану.

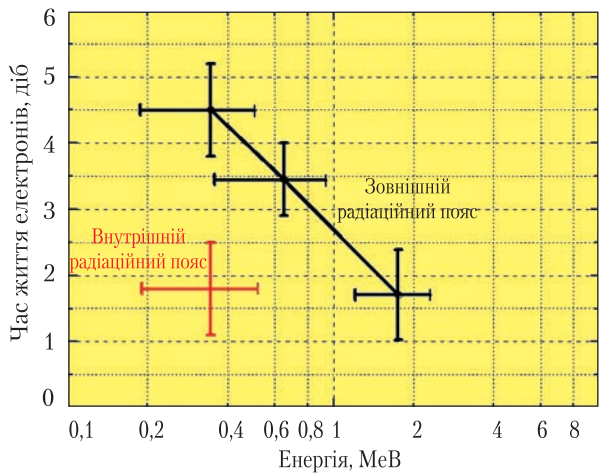

**Рис. 13.** Емпіричні часи життя електронів у радіаційних поясах залежно від їх енергії (за даними приладу СТЕП-Ф)

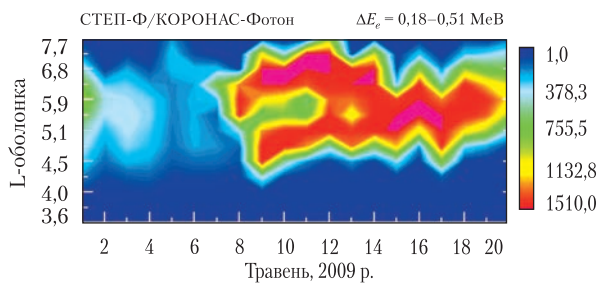

**Рис. 14.** Розщеплення зовнішнього електронного радіаційного поясу за даними приладу СТЕП-Ф у діапазоні енергій електронів $E_e = 0{,}18{-}0{,}51$ МеВ на фазі поновлення геомагнітної бурі 6–8 травня 2009 р.

*Розщеплення зовнішнього радіаційного поясу.* З попередніх супутникових експериментів відомо, що на фазі поновлення геомагнітної бурі відбувається прискорення заряджених частинок і збільшення щільності їх потоків у зовнішньому радіаційному поясі. В окремих випадках, під час потужних геомагнітних штормів, щілина між внутрішнім і зовнішнім поясами заповнюється частинками настільки, що обидва пояси з'єднуються між собою і виглядають як один широкий у радіальному напрямку пояс. Прилад СТЕП-Ф продемонстрував наявність ще одного феномену — роз-щеплення зовнішнього поясу на два близьких (за дрейфовими L-оболонками) пояси на короткий проміжок часу (рис. 14). Час існування розщепленого поясу збігається з часом поновлення магнітної суббурі; просторова відстань між сусідніми дрейфовими оболонками з підвищеними потоками частинок залежить, імовірно, від інтенсивності бурі. Достеменно поки відомо, що розщеплення не пов'язане із заповненням проміжку між радіаційними поясами Ван Алена, просторова відстань між сусідніми максимумами зовнішньої зони скорочується зі зростанням енергії електронів. Пояснення причин розщеплення поки що дискутується: це може бути відоме здавна явище розщеплення дрейфових оболонок або ж інжекція вузькоспрямованих пучків електронів з міжпланетного простору на різні оболонки після азимутального дрейфу навколо Землі. У будь-якому разі це явище є досить новим і потребує подальшого вивчення.

## Оприлюднення результатів наукового експерименту з використанням приладу СТЕП-Ф

Етапи процесу створення супутникового приладу СТЕП-Ф та наукові результати, отримані після обробки експериментальних даних, неодноразово висвітлювалися у ЗМІ та на офіційних веб-сайтах Державного космічного агентства України, НАН України, МОН України, УНТЦ, ІАФ МІФІ, організацій — розробників складових приладу.

Результати моделювань, тестувань, випробувань детекторів і електронних модулів приладу було представлено на численних вітчизняних та міжнародних конференціях, робочих нарадах, семінарах, опубліковано в українських і зарубіжних наукових журналах. Зокрема, матеріали щодо концепції та опису приладу доповідалися на регулярних симпозіумах COSPAR, міжнародних конференціях з фізики космічних променів, європейських астрономічних конференціях JENAM, конференціях Європейського геофізичного товариства (EGU), щорічних українських конференціях з космічних досліджень тощо. Наукові результати, отримані з використанням інформації з приладу СТЕП-Ф, були також репрезентовані на багатьох конференціях і семінарах в Україні, Росії, Фінляндії, Польщі, США; публікувалися в міжнародних наукових журналах. Загалом було опубліковано 56 статей і тез доповідей з опису супутникового приладу СТЕП-Ф, його тестувань та наукових результатів, отриманих за його допомогою. Найбільш вагомі публікації наведено у пристатейному списку літератури [1—17].

Українсько-польське співробітництво з обробки даних з приладу СТЕП-Ф і сонячного рентгенівського спектрофотометра SphinX виявилося дуже плідним — у 2014 р. авторському колективу Радіоастрономічного інституту НАН України і Центру космічних досліджень ПАН було присуджено премію Національної академії наук України і Польської академії наук за видатні результати, одержані під час виконання циклу досліджень «Динамічна структура високоенергетичних частинок у навколоземному просторі» за даними польського фотометра SphinX і українського телескопа електронів СТЕП-Ф на навколоземному супутнику «КОРОНАС-Фотон» і підготовки до спільного космічного експерименту на борту міжпланетного зонда «Інтергеліозонд».

Групу розробників приладу СТЕП-Ф нагороджено відзнаками Кабінету Міністрів України, Державного космічного агентства України, Національної академії наук України, Міністерства освіти і науки України, Харківської обласної ради, Харківського міського голови та ін.

Отже, створення супутникового приладу СТЕП-Ф є результатом плідної співпраці установ МОН України, НАН України і Державного космічного агентства України та конструктивного багаторічного наукового співробітництва вчених Харківського національного університету ім. В.Н. Каразіна та Інституту астрофізики МІФІ. Через відсутність українських наукових космічних апаратів та хронічне недофінансування космічної галузі цей космічний експеримент вимушено було виконано на закордонному супутнику, що стало чи не єдиним українським тривалим дослідницьким експериментом у космосі, принаймні за останнє десятиліття. З використанням приладу СТЕП-Ф було накопичено і оброблено неоцінені наукові дані про потоки частинок високих енергій навколо нашої планети в період глибокого мінімуму сонячної активності, що поклало початок довготривалій науковій співпраці з ученими і розробниками космічної апаратури Центру космічних досліджень Польської академії наук.

REFERENCES
[СПИСОК ЛІТЕРАТУРИ]

1. Zalyubovsky I.I., Dudnik A.V., Kotov Yu.D., Yurov V.N. The International project PHOTON for the complex study of solar activity and Sun-Earth connections. *Bulletin of the Russian Academy of Sciences: Physics*. 1997. **61**(6): 1173.
   [Залюбовский И.И., Дудник А.В., Котов Ю.Д., Юров В.Н. Международный проект «Фотон» для комплексного изучения солнечной активности и солнечно-земных связей. *Известия РАН. Серия физическая*. 1997. Т. 61. № 6. С. 1173—1176.]

2. Dudnik O.V., Zalyubovsky I.I. Scientific tasks of international space experiment "CORONAS-PHOTON". *Kosm. Nauka Tehnol.* 2000. **6**(2): 3.
   [Дудник А.В., Залюбовский И.И. Научные задачи международного космического эксперимента КОРОНАС-ФОТОН. *Космічна наука і технологія*. 2000. Т. 6, № 2. С. 3—12.]

3. Dudnik O.V., Malykhina T.V. The computer simulation of deposited energies and stopping ranges of particles in the STEP spectrometer of the Warning space project. *Kosm. Nauka Tehnol.* 2003. **9**(1): 15.
   [Дудник А.В., Малыхина Т.В. Компьютерное моделирование поглощенной энергии и пробегов частиц в спектрометре СТЭП космического проекта «Попереждение». *Космічна наука і технологія*. 2003. Т. 9. № 1. С. 15—21.]

4. Dudnik O.V., Goka T., Matsumoto H., Fujii M., Persikov V.K., Malykhina T.V. Computer simulation and calibration of the charge particle spectrometer—telescope STEP-F. *Adv. Space Res.* 2003. **32**(11): 2367.

5. Dudnik O., Goka T., Matsumoto H., Fujii M., Persikov V., Malykhina T., Kato H. Accelerator test of charge particle detectors for a satellite instrument STEP-F. *RIKEN Accelerator Progress Report* (Japan). 2003. **37**: 168.

6. Dudnik O.V., Persikov V.K., Zalyubovskiy I.I., Timakova T.G., Kurbatov E.V., Kotov Yu.D., Yurov V.N. High-sensitivity STEP-F spectrometer—telescope for high-energy particles of the CORONAS-PHOTON satellite experiment. *Solar System Research*. 2011. **45**(3): 212.
   [Дудник А.В., Персиков В.К., Залюбовский И.И. и др. Высокочувствительный спектрометр-телескоп высокоэнергетических частиц СТЭП-Ф спутникового эксперимента КОРОНАС-ФОТОН. *Астрономический вестник*. 2011. Т. 45. № 3. С. 219—227.]

7. Dudnik O.V. Investigation of the Earth's radiation belts in May, 2009, at the low orbit satellite with the STEP-F instrument. *Kosm. Nauka Tehnol.* 2010. **16**(5): 12.
   [Дудник А.В. Исследование радиационных поясов Земли в мае 2009 года на низкоорбитальном спутнике с помощью прибора СТЭП-Ф. *Космічна наука і технологія*. 2010. Т. 16. № 5. С. 12—28.]

8. Dudnik O.V. Variations of electron fluxes in the Earth radiation belts in May, 2009 due to "STEP-F" device observations. *Journal of Kharkov National University. Physical series.* 2010. **916**(3): 56.
   [Дудник А.В. Вариации потоков электронов в радиационных поясах Земли в мае 2009 года по наблюдениям с помощью прибора «СТЭП-Ф». *Вісник Харківського національного університету ім. В.Н. Каразіна, серія фізична «Ядра, частинки, поля»*. 2010. Т. 916. Вип. 3. С. 56—66.]

9. Dudnik O.V., Podgorski P., Sylwester J., Gburek S., Kowalinski M., Siarkowski M., Plocieniak S., Bakala J. Investigation of electron belts in the Earth's magnetosphere with the help of X-ray spectrophotometer SphinX and satellite telescope of electrons and protons STEP-F: preliminary results. *Kosm. Nauka Tehnol.* 2011. **17**(4): 14.
   [Дудник А.В., Подгурски П., Сильвестер Я. и др. Исследования электронных поясов в земной магнитосфере с помощью рентгеновского спектрофотометра SphinX и спутникового телескопа электронов и протонов СТЭП-Ф: предварительные результаты. *Космічна наука і технологія*. 2011. Т. 17. № 4. С. 14—25.]

10. Dudnik O.V., Podgorski P., Sylwester J., Gburek S., Kowalinski M., Siarkowski M., Plocieniak S., Bakala J. X-Ray spectrophotometer SphinX and particle spectrometer STEP-F of the satellite experiment CORONAS-PHOTON. preliminary results of the joint data analysis. *Solar System Research*. 2012. **46**(2): 160.
   [Дудник А.В., Подгурски П., Сильвестер Я. и др. Рентгеновский спектрофотометр SphinX и спектрометр частиц СТЭП-Ф спутникового эксперимента КОРОНАС-ФОТОН — предварительные результаты совместного анализа данных. *Астрономический вестник*. 2012. Т. 46. № 2. С. 173—183.]

11. Dudnik O.V. Dynamics of the Earth's radiation belt electrons in May, 2009, on the base of the STEP-F instrument observations. In: *Results of space experiment "CORONAS-PHOTON"*: Proc. Workshop (May 19—21, 2010, Tarusa, Russia). (Moscow. IKI RAN, 2012). P. 103—125.
   [Дудник А.В. Динамика электронов радиационных поясов Земли в мае 2009 г. по наблюдениям с помощью аппаратуры СТЭП-Ф. В кн.: *Результаты космического эксперимента «КОРОНАС-ФОТОН»*: сб. тр. семинара (19—21 мая 2010, Таруса, Россия). М.: ИКИ РАН, 2012. С. 103-125.]

12. Podgorski P., Dudnik O.V., Sylwester J., Gburek S., Kowalinski M., Siarkowski M., Plocieniak S., Bakala J. Joint analysis of SphinX and STEP-F instruments data on magnetospheric electron flux dynamics at low Earth orbit. *Proc. 39th COSPAR Scientific Assembly* (July 14—22, 2012, Mysore, India). P. 112.

13. Dudnik O. Unexpected behavior of subrelativistic electron fluxes under Earth radiation belts. In: HEPPA/SOLARIS-2012: 4th Int. workshop (October 9—12, 2012, Boulder, Colorado, USA). P. 15.

14. Dudnik O., Sylwester J., Podgorsi P., Gburek S. Radiation belts of the Earth: overview, methods of investigations, recent observations on the CORONAS-Photon spacecraft. In: *Progress on EUV&X-ray spectroscopy and imaging*: Proc. Int. Conf. (November 20—22, 2012, Wroclaw, Poland).

15. Dudnik O.V., Sylwester J., Podgorski P. Properties of magnetospheric high energy particles based on analysis of data from STEP-F and SphinX instruments aboard the "CORONAS—PHOTON" satellite. In: *Space Research in Ukraine, 2012—2014. The Report to the COSPAR*. Fedorov O.P. (ed.). (Kyiv: Akademperiodyka, 2014). P. 53-61.
   [Дудник А.В., Сильвестер Я., Подгурски П. Исследования частиц высоких энергий на низкоорбитальном спутнике «КОРОНАС-ФОТОН» по данным приборов СТЭП-Ф и SphinX. В кн.: *Космічні дослідження в Україні, 2012—2014. Звіт до COSPAR*. За ред. О.П. Федорова. К.: Академперіодика, 2014. С. 56—64.]

16. Dudnik O.V., Podgorski P., Sylwester J. New perspectives to study the splitting of drift shells at the outer magnetosphere by using STEP-F and SphinX instruments on board the CORONAS-Photon satellite. In: *Progress on EUV&X-ray spectroscopy and imaging II*: Proc. Int. Conf. (November 17—19, 2015, Wroclaw, Poland). P. 8.

17. Dudnik O.V., Podgorski P., Sylwester J.Combined study of radiation belts by the satellite particle telescope STEP-F and solar soft X-ray photometer SphinX during recent deep minimum of solar activity. In: *Proc. 23 National Solar Physics Meeting* (May 30 — June 3, 2016, Liptovský Mikuláš, Slovakia). P. 1—6.

*O.V. Dudnik*

Institute of Radio Astronomy of the National Academy of Sciences of Ukraine (Kharkiv)
http://orcid.org/0000-0002-5127-5843

SATELLITE TELESCOPE OF ELECTRONS AND PROTONS STEP-F
OF THE SPACE SCIENTIFIC PROJECT CORONAS-PHOTON

The Ukrainian satellite telescope of electrons and protons STEP-F provided scientific experiment in the space on the board of low Earth orbit spacecraft CORONAS-Photon in 2009. In article it is described the background, stages of development, manufacturing and tests of weight-dimensional, breadboard, technological, flight and auxiliary models of the instrument, and of the control and test suites for the STEP-F. Principles of operation, design, technical and scientific characteristics of the instrument have been described. Stages of adjustment, graduating and of autonomous, docking, complex, ground pre-flight and flight tests have been shown. Scientific results obtained by the STEP-F instrument in time of deep minimum of solar activity demonstrate detected new features in high energy charge particle distributions inside radiation belts of the Earth, in the region of Brazil magnetic anomaly, and outside of referred volumes.

*Keywords*: Earth's magnetosphere, satellite instrument, scintillation detector, silicon matrix, charge particles, radiation belt, solar activity, Brazil magnetic anomaly.